\documentclass[12pt]{article}
\usepackage{setspace}
\usepackage{geometry}
\usepackage{multirow}
\usepackage{xcolor}
\usepackage{amssymb}

\usepackage{graphicx}
\graphicspath{{../credible_reviews_paper/Figures/}{./Figures/}}

\usepackage{amsmath,amssymb}
\usepackage{caption}
\usepackage{subcaption}
\usepackage{booktabs}
\usepackage{float}
{\newgeometry{left=1in,right=1in,top=1in,bottom=1in}
\usepackage[round]{natbib}

\begin{document}




\title{Understanding and Overcoming Biases in Customer Reviews}

\author{
  Georgios Askalidis\\
  {\it Northwestern University}
  \and
  Edward C. Malthouse\\
  {\it Northwestern University}
}
\maketitle
\begin{abstract}
Our paper contributes to the literature recommending approaches to make online reviews more credible and representative. We analyze data from four diverse major online retailers and find that verified customers who are prompted (by an email) to write a review, submit, on average, up to 0.5 star higher ratings than self-motivated web reviewers. Moreover, these email-prompted reviews remain stable over time, whereas web reviews exhibit a downward trend. This finding provides support for the existence of social influence and selection biases during the submission of a web review, when social signals are being displayed. In contrast, no information about the current state of the reviews is displayed in the email promptings. Moreover, we find that when a retailer decides to start sending email promptings, the existing population of web reviewers is unaffected both in their volume as well as the characteristics of their submitted reviews. We explore how our combined findings can suggest ways to mitigate various biases that govern online review submissions and help practitioners provide more credible, representative and higher ratings to their customers.


\end{abstract}


\section{Introduction}
\par `Word of Mouth' (WOM), defined as an informal communication between private parties concerning the evaluation of goods and services \citep{westbrook1987product, singh1988consumer, fornell1982two, dichter1966word}, has been part of human behavior for a long time. With the rise of the Internet, WOM has evolved and changed. Even though `traditional' WOM will not be eliminated anytime soon (e.g., think of friends and family telling us about the latest shows they've watched on TV), `electronic' word of mouth (eWOM) \citep{hennig2004electronic} offers the significant advantage that users are no longer forced to rely on scattered signals from their immediate social network to be informed about the quality of a product, but instead can access reviews from all over the world in an organized and on-demand way.

\par Indeed, reviews are being collected, aggregated and displayed to consumers in an easy-to-digest format in all types of settings: all of the top-10 U.S.\ online retailers (as well as most of the biggest retailers in the rest of the world, such as Alibaba) collect and display user reviews for their products. The same is true for all the major digital stores. Furthermore, companies like Yelp, Facebook, Google, IMDb and Rotten Tomatoes provide platforms for users to submit reviews that are in-turn aggregated and displayed to other users. User reviews are also being used to build trust between customers in decentralized marketplaces like eBay, Airbnb and Uber. This trust between users is a cornerstone for the success of any such marketplace, where customers interact and make financial transactions with strangers.
\par For online shoppers, reviews are not just an option anymore but an expectation. A recent survey\footnote{http://www.powerreviews.com/blog/survey-confirms-the-value-of-reviews/} found that 30\% of shoppers (under the age of 45) consult reviews for every purchase they make, while 86\% say that reviews are essential in making purchase decisions. In fact, after price, reviews are the factor with the most impact on purchases. 
\par Apart from the widespread {\it adoption} of online reviews, an extensive literature has showcased the economic {\it importance} of positive reviews. A 1-star increase in the Yelp ratings of a restaurant can cause a 5--9\% increase in revenue~\citep{luca2011reviews}, and an extra half-star can help a restaurant sell out its reservations 50\% more frequently~\citep{anderson2012learning}. Positive correlations between ratings and sales have been found for products on Amazon~\citep{chevalier2006effect}, for new products~\citep{cui2012effect}, for  movies \citep{dellarocas2005using}\footnote{\cite{duan2008online} found significant correlation between a movie's box office revenue and the {\it volume} of online user reviews, but not with the ratings} and for apps in Google's mobile app store,~\citep{engstrom2014demand}. On two-sided marketplaces such as eBay, an extensive literature has found that positive user feedback leads to economic benefits~\citep{cabral2010dynamics, houser2006reputation, resnick2006value}. 
\par Besides the widespread adoption and demonstrated economic significance of online user reviews, another line of research has examined the biases that govern the submission of online user reviews. {\it Social influence bias}, which, roughly speaking, is when a user's opinion is influenced by the opinions of other users, is one of the main ones studied in the literature. For example,~\cite{muchnik2013social} showed that an arbitrary positive vote on a comment submitted to a news aggregator website created accumulating positive herding that increased final ratings by 25\% on average.~\cite{salganik2006experimental} created an artificial music market where participants downloaded previously unknown songs, either with or without information about the previous participants' actions, and found that the display of social signals increases the inequality and unpredictability of success.
\par In addition to social influence, another bias that has been studied in the literature is the {\it selection bias} which, roughly speaking, is when the set of users that submit a review is not representative of the entire purchasing population. For example, \cite{hu2009overcoming} have demonstrated that review distributions on online platforms tend to be bi-modular, suggesting that extremely satisfied and extremely dissatisfied customers are more likely to submit a review. Furthermore,~\cite{li2008self} and~\cite{godes2012sequential} have shown that online reviews exhibit temporal trends, indicating that users who submit a review later in a product's life cycle are generally different than users that review earlier. Moreover, the propensity of a user to review can be a function not only of their opinion about the product, but also of the current state of the reviews \citep{nagle2014online}.

\par With online reviews being omnipresent, economically influential and biased, the success of establishments, products or agents in a two-sided marketplace can be decided by factors other than their true quality. Hence there is a need to understand the biases that govern online reviews and suggest ways to fix them. Our paper is contributing to this literature.
\par We ask and explore two main questions. First, how do different populations differ when they write reviews for the same set of products? We examine two populations: (1) users that are {\it self-motivated} to write a review, i.e., users that, after their purchase, visited a retailer's webpage and did all the necessary steps to submit a review, and (2) {\it prompted} reviewers, i.e., users that submitted a review after receiving an email from (or on behalf of) a retailer soliciting a review for a product they recently bought. Throughout this work, we will refer to self-motivated reviews as {\it web reviews} indicating the fact that they came through the web and to email-prompted reviews as {\it email reviews}. Accordingly, we will refer to the author of a web (email) review as a {\it web (email) reviewer}. We find that email reviews are significantly and substantially more positive than web reviews, indicating that dissatisfied customers are more likely to be self-motivated to write a review. This is a finding that is in line with (and perhaps an eWOM version of) \cite{anderson1998customer}, that found that dissatisfied customers engage in greater WOM than satisfied ones. Moreover, we find that email reviews are stable over time while web reviews exhibit a downward trend, indicating that the display of various social signals throughout the process of a web review submission induces selection and perhaps social influence biases. 
\par Since, for some retailers, soliciting reviews using emails is a relatively recent phenomenon, we are interested in understanding how the introduction of these email promptings affected the entire review ecosystem and the existing reviewing population in particular. Hence, our second question has two parts. How did the reviewing population and their submitted reviews change? And, in particular, how did the self-motivated reviewing population and the reviews they submit change as a result of the introduction of email prompts. Even though we find an overall increase in volume and star-rating average, we find no evidence of disturbances in the self-motivated population or in their submitted reviews. This indicates that sending email prompts taps into an entirely new segment of the purchasing population without disturbing the population that is already reviewing, making the new set of reviews more representative. Since the new population of email reviewers are all verified buyers (email promptings are sent only to verified buyers), the reviews overall become also more credible. And finally, since email reviews carry higher ratings than web reviews, the new set of reviews becomes more positive.
\par Our dataset  is comprised of the entire review history of four major online retailers in different categories that, between the four of them, sell a wide variety of electronics, appliances, bedding, kitchen, jewelry, personal care and health products. Each datapoint represents a submitted review and carries, amongst other, the following information: review rating, review text, date submitted, product id, number of `Helpful' votes, number of `Not Helpful' votes, and a source. The source can take two values: `web' or 'email', indicating if the review is a web or email review, as defined above.

\section{Differences Between Self-Motivated and Prompted Reviews}\label{differences}
\par In this section we examine the differences that email and web reviews exhibit with respect to some key metrics, including review rating and volume. 
\par Our dataset consists of 238,809 reviews for 27,574 unique products, across four major online retailers. For each review we know the {\it review rating}, which is an integer between 1 and 5 indicating the number of stars that the user submitted for the product and {\it review text}, which is the actual text of the review accompanying the rating.
\par In order to study temporal trends, we calculate each review's {\it arrival rank}. This is an integer indicating the chronological order in which a review arrived, amongst all other reviews for the same product. Note that the arrival rank doesn't take into account the actual time a review was submitted, but only the relative order amongst all other reviews for the same product. 
\par Our dataset also captures the votes (`Helpful' or `Not Helpful') that users submitted regarding the helpfulness of an existing review. Using this data, we compute, for each review, the variables {\it helpful votes}, indicating how many users voted on the helpfulness of that review, and {\it helpful score} indicating the percentage of those votes that were positive. Note that no registration or purchase is required for casting a vote on the helpfulness of an existing review.
\par
\begin{figure}[h!]
\centering
\begin{subfigure}{.45\textwidth}
\centering
\includegraphics[width=\textwidth]{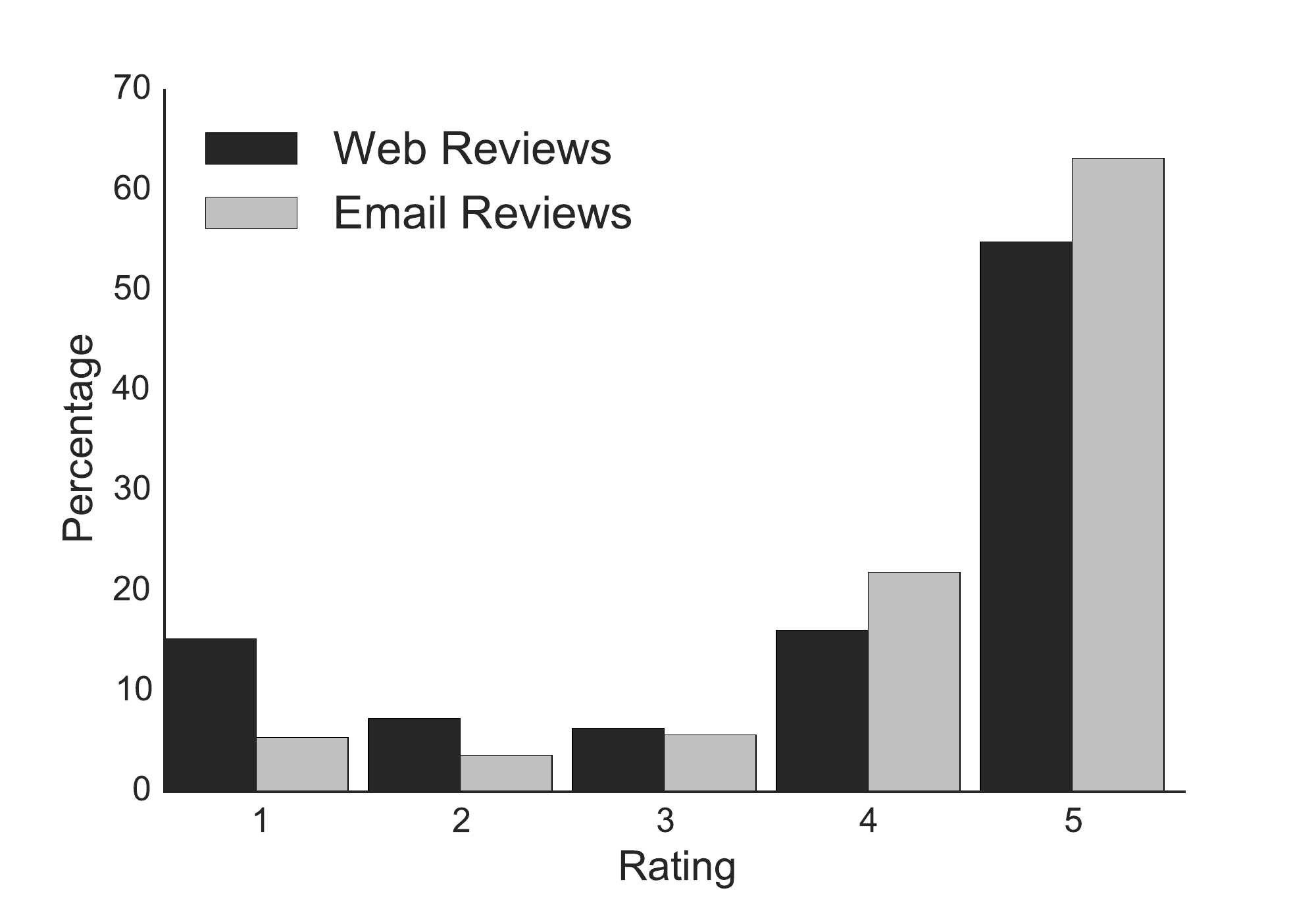}
\caption{Rating Distributions}
\label{rating_dist}
\end{subfigure}
\quad
\begin{subfigure}{.45\textwidth}
\centering
\includegraphics[width=\textwidth]{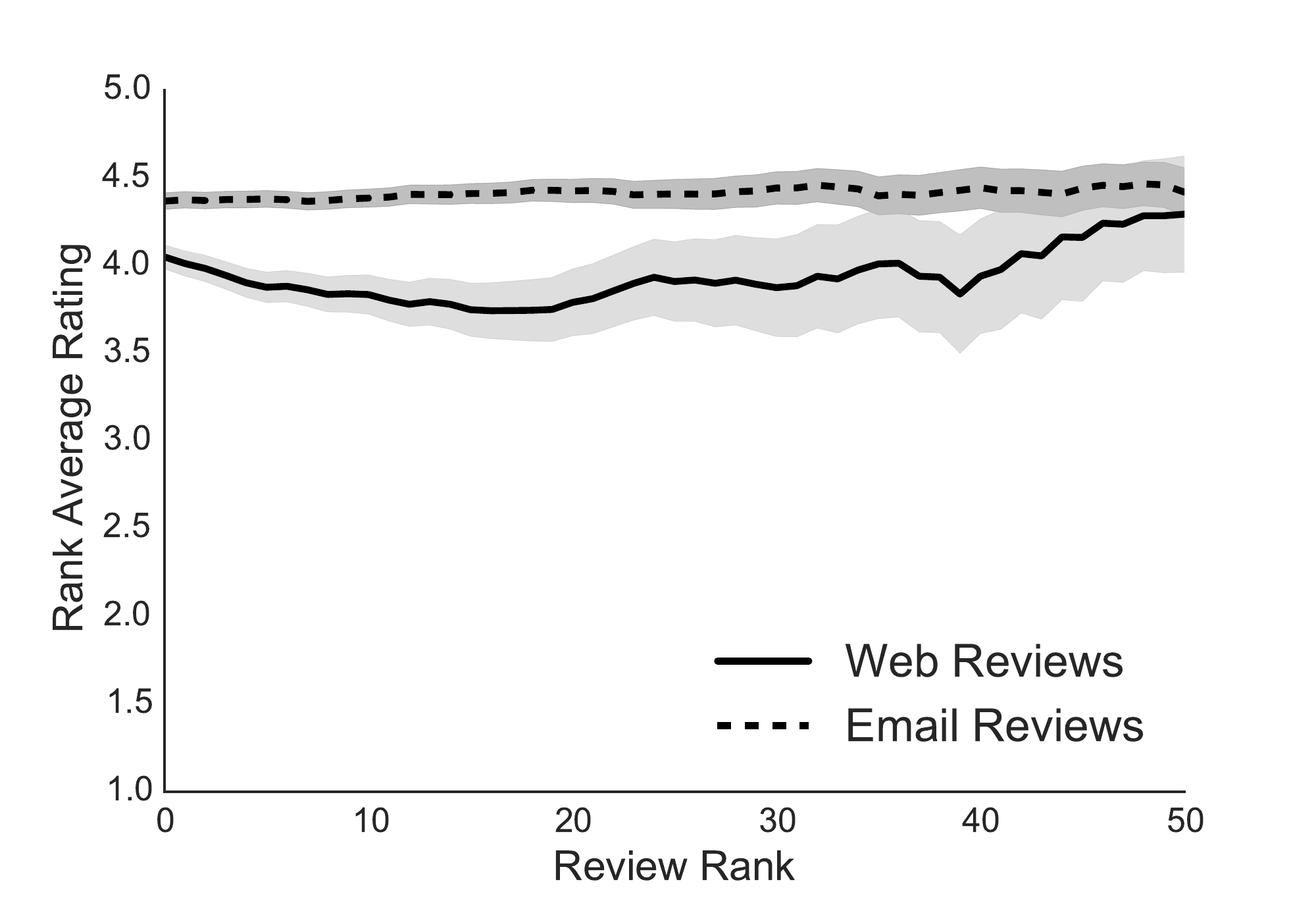}
\caption{Rating Evolution by Arrival Rank, with 95\% confidence bands}
\label{rating_evol_rank}
\end{subfigure}
\caption{Rating Distribution and Evolution for Web and Email Reviews.}
\label{email_v_web_evol}
\end{figure}

  ~\cite{anderson1998customer} found that dissatisfied customers are more likely to be vocal about their dissatisfaction than satisfied customers about their satisfaction, and we expect this phenomenon to induce a selection bias, where self-motivated reviewers are more likely to be dissatisfied. Hence, we expect to see a larger percentage of lower ratings coming from web reviews than email reviews. Indeed, we find that the average web rating is 3.88 compared to 4.3 for email reviews. A look at the distributions of the two sets of reviews, shown in Figure \ref{rating_dist}, provides further confirmation that web reviews tend to be more negative than email reviews. 
We notice that email reviews have a higher percentage of 4- and 5-star ratings, a roughly equal percentage of 3-star ratings and substantially lower percentage of 1- and 2-star ratings.

 Hence, the distribution of web reviews is `J-shaped' (see e.g., \cite{hu2009overcoming}) which is frequently observed in online user platforms that are populated mainly by self-motivated reviews (such as Amazon and Yelp).
\par Each email review lives in its own silo: there are no social signals on the email promptings sent, and if a reviewer decides to follow the link and submit a review as a result of the prompting they are taken to an isolated page where no information about the current state of the reviews is displayed. In contrast, web reviewers observe social signals about the current state of the reviews throughout the entire reviewing process. Hence, not only the review that a user submits can be influenced by the existing reviews (i.e., social influence bias) but even the decision of a user to submit a review can be influenced by the current state of the reviews (i.e., selection bias).  Indeed, \cite{godes2012sequential} observed a downward slope for reviews on Amazon.com (a platform that is comprised mainly by self-motivated reviews). Hence, we would expect to see a similar temporal trend from the web reviews in our dataset but not from the email reviews. Figure~\ref{rating_evol_rank} displays the evolution of average ratings by the review's arrival rank with 95\% confidence bands. As expected, the plot suggests that email reviews are stable over time (i.e., the 20$^\mathrm{th}$ email review for a product is, on average, equal to the 1$^\mathrm{st}$ email review for that product), while web reviews display a downward temporal trend.

\par We also expect to see differences in the {\it text} that is submitted by web and email reviews. A large body of work has demonstrated that intrinsic motivation is a strong predictor of high quality work, see e.g., \cite{cerasoli2014intrinsic} for a survey of that literature. In this paper we focus on the review text length as an approximation to its quality, and hence we expect to see web reviews to be longer. Other measures for the quality of the review text could be explored in future work. Indeed, we find that web reviews have on average 300 characters compared to 160 for email reviews.

\par When products have numerous reviews, users may be selecting just a few to read. Various platforms, in an effort to help customers identify influential or high quality reviews, allow browsing users to provide feedback on the existing reviews. This feedback is usually in the form of a positive or negative vote, indicating if the review was {\it helpful} to the reading user or not. Many platforms allow customers to sort the existing reviews according to their helpfulness and some take it one step further by making this display ordering their default one. Hence, we expect these helpful reviews to be disproportionally influential in the purchase decisions of browsing customers, and we seek to understand better what are the factors that make them being perceived as helpful. Previous literature has shown a positive correlation between lower ratings and higher perceived helpfulness \citep{bakhshi2014if}. Since, in general, web reviews carry lower ratings than email reviews we would expect to see web reviews to have a higher helpful score (i.e., number of positive votes divided by the number of all votes). Interestingly, we observe no substantial difference, with the helpful score being around 83\% for both sets of reviews. Where we do see a difference, is on helpful votes, i.e., the number of votes a review received. Web reviews receive 1.2 votes on average while email reviews receive 0.8. This is despite web reviews generally arriving later than email reviews. This higher number of votes could be explained by the larger text that web reviews have and which might indicate a more in-depth analysis of the product. In fact, recent work has shown a positive correlation between a review's text length and it's readability, \cite{salehan2016predicting}

\subsection{Econometric Model}\label{econ_model_1}
\par Following our exploratory analysis, we now turn to an econometric model to provide statistical tests for our descriptive results. Our general econometric model is as follows,
\begin{equation}\label{model_1}
y= \alpha_0 + \beta_1\mathrm{web} +\beta_2\mathrm{rank} + \beta_3\mathrm{web}\cdot\mathrm{rank}+e,
\end{equation}
where web is a binary variable indicating a web review, and rank is an integer variable indicating the arrival rank of the review. The coefficient of the rank variable will detect any temporal trends the reviews may exhibit. We add the interaction variable web$\cdot$rank to detect any differences in the temporal trends that each set of reviews exhibit. The error term is $e$.
\par We start by estimating Model~\ref{model_1} with the review rating being the dependent variable. This will show if the differences we observe in the average rating and temporal trends between web and email reviews are statistically significant. We also estimate Model \ref{model_1} with the review length as the dependent variable, in order to explore any differences between the text of the reviews submitted by web and email reviews. Finally,  we seek to understand if there are differences between how many helpful votes web and email reviews receive, and if one set of reviews is generally perceived, by the users, as more helpful. Hence, we estimate Model \ref{model_1} with respect to the helpful votes and helpful score metrics as well.
\par Table \ref{regressions_1} summarizes the results of the estimations of Model \ref{model_1}.

\begin{table*}[h!]
\begin{center}
\vspace{0.5cm}
\begin{tabular}{l c c c c  c} 
 \toprule
		{\it \small Dependent Variable}						&Intercept 				&web 				& rank 	&  web$\cdot$rank\\[6pt]
\midrule
\multirow{2}{*}{Review Rating} & $4.366^{***}$			&$-0.374^{***}$			&$-2.5\cdot 10^{-5}$			& $-0.0125^{***}$	\\[6pt]	
								&(0.008)				&(0.012)				&(0)			&(0.001)			&			\\[6pt]

\midrule

\multirow{2}{*}{\small Log Review Length} 	&   $4.668^{***}$	& $0.7608^{***}$	&$-0.0036^{***}$ 		& 0.0018\\[6pt]	
											&(0.006)			&(0.01)				&(0)					&(0.001)		&\\[6pt]

\midrule

\multirow{2}{*}{Helpful Votes} 	& $1.187^{***}$		& 	$0.67^{***}$			&$-0.03^{***}$ 		& $-0.011^{***}$	\\[6pt]
								&(0.012)					&(0.02)					&(0.001)				&(0.001)		&\\[6pt]
\midrule

\multirow{2}{*}{Helpful Score} 	&  $0.83^{***}$			& $0.011^{*}$			&$-0.0013^{***}$ 		& $-0.0014^{***}$	\\[6pt]
									&(0.003)					&(0.005)			&(0.000)				&(0.000) &	\\[6pt]


\bottomrule
\end{tabular}
\end{center}
Values in parentheses are standard errors.\\
\small{$^{*}:p<0.05$, $^{**}: p<0.01$, $^{***}:p<0.001$}
\caption{Quantitative and qualitative differences between web and email reviews}
\label{regressions_1}
\end{table*}

\subsection{Results}
We discuss here the results from the estimations of Model \ref{model_1} and how they confirm the exploratory results we presented above.
\paragraph{Rating}
\par The first row of Table~\ref{regressions_1} shows the result of the Model~\ref{model_1} estimation with review rating as the dependent variable, and it shows a highly significant, substantial and negative coefficient for the web variable. This result confirms our exploratory analysis findings, shown in Figure~\ref{rating_dist}, that email reviews are 0.37 higher than web reviews.
\par Furthermore, the coefficient for rank is not statistically significantly different from zero, indicating that email reviews do not display any temporal trends whereas the coefficient for web$\cdot$rank is negative and highly significant, indicating a downward slope for web reviews. This finding agrees with the exploratory analysis finding shown in Figure \ref{rating_evol_rank}, and with previous literature that has focused on self-motivated reviews \citep{godes2012sequential}. In fact, in further agreement between the behavior of the web reviews in our dataset and the (mainly self-motivated) reviews of \cite{godes2012sequential}, we observe a similar downward trend even if we order the reviews by their arrived  {\it time}, i.e., how many days in the life cycle of the product they were submitted. Email reviews are stable over time, even with respect to this metric.
\par These differences between the temporal trends of web and email reviews provide support for the existence of selection and social influence biases when social signals are being displayed during the review process (as it happens with web reviews and doesn't happen with email reviews).
\paragraph{Review Text} The estimation of Model~\ref{model_1} with the logarithm of the review length as a dependent variable, shown in the second row of Table~\ref{regressions_1}, shows that indeed web reviews have statistically significant and substantially larger text. Since web reviewers are self-motivated, this finding may be related to an extensive literature that has shown the intrinsic motivation produce higher quality results than external incentives (see e.g., \cite{cerasoli2014intrinsic} for a survey). Furthermore, the estimation shows no temporal trend for the review length of web reviews and a significant but very weak downward trend for the review length of email reviews.


\paragraph{Helpfulness} Finally, we turn our attention to helpfulness. Note that a browsing user cannot distinguish if a review comes from the web or email. They only see a`Verified Buyer' sticker under each review that comes from such a customer. All email reviews carry that sticker and almost none of the web reviews do, although review readers do not know that verified buyers are nearly synonymous with email reviews.
\par Our exploratory analysis showed that web reviews receive generally more votes (positive or negative) regarding their helpfulness and the third row of Table \ref{regressions_1} confirms this finding as highly statistically significant. This is related to recent work by \cite{salehan2016predicting} who found a positive correlation between the length of a review and it's readership. Moreover, as reviews that are in the system longer have more time to accumulate votes, one would expect the number of votes a review receives to decline with respect to its arrival rank. Indeed, the highly statistically significant and negative values for the coefficients of rank and web$\cdot$rank, shown in Table \ref{regressions_1}, confirm that expectation for both email and web reviews.
\par The number of votes a review receives, however, is perhaps not as important as the percentage of those votes that are positive. As we explained earlier, we define the {\it helpful score} of a review to be the number of `helpful' votes it received divided by the total number of votes it received. In our exploratory analysis we found that both web and email reviews have a helpful score of around 83\%. Even though the coefficient for web, shown in the fourth row of Table~\ref{regressions_1}, is positive and significant, indicating that web reviews have a higher helpful score than email reviews, its coefficient is fairly small (0.01) compared to the intercept (0.83). 
\par The negative and highly statistically significant coefficient for rank and web$\cdot$rank indicate that reviews with higher arrival ranks (i.e., reviews that are submitted later in a product's life cycle) have, on average, lower helpful scores. This can be explained by the fact that for all four of the retailers in our dataset, the default sorting of the reviews is by their helpful score. Hence, a rich-get-richer effect takes place with reviews that are perceived as helpful being displayed to more users and being more likely to receive further positive helpful votes.


\section{The Effect of the Introduction of Email Reviews}
\par Some of the retailers in our dataset operated for many years, since their creation, without sending email promptings to their customers. Hence, we see the opportunity to approach the introduction of the email promptings as a natural experiment, and explore the effect it had on the retailer's review ecosystem.
\par In order to approximate the exact date that a retailer started sending out email promptings to their customers, we observe the first email review that was submitted in that retailer's platform. We then keep all the reviews that were submitted in the three months before the email review introduction and the three months after. We consider this set of reviews to be the `treatment' year, in contrast to another set of reviews we get for control. Our control set of reviews consists of those submitted in the same six-month period exactly one year before the treatment. Hence, if the email reviews were introduced for a retailer on June 1, 2012 our treatment group of reviews is comprised of those submitted in the period March 1 -- September 1, 2012 while our control group consists of those submitted in the period March 1 --  September 1, 2011. We then use a difference-in-differences approach to understand any effect the email introduction had. We use the control group of reviews to control for temporal and other trends.
\par
\begin{figure}[h!]
\begin{subfigure}{.45\textwidth}
\centering
\includegraphics[width=\textwidth]{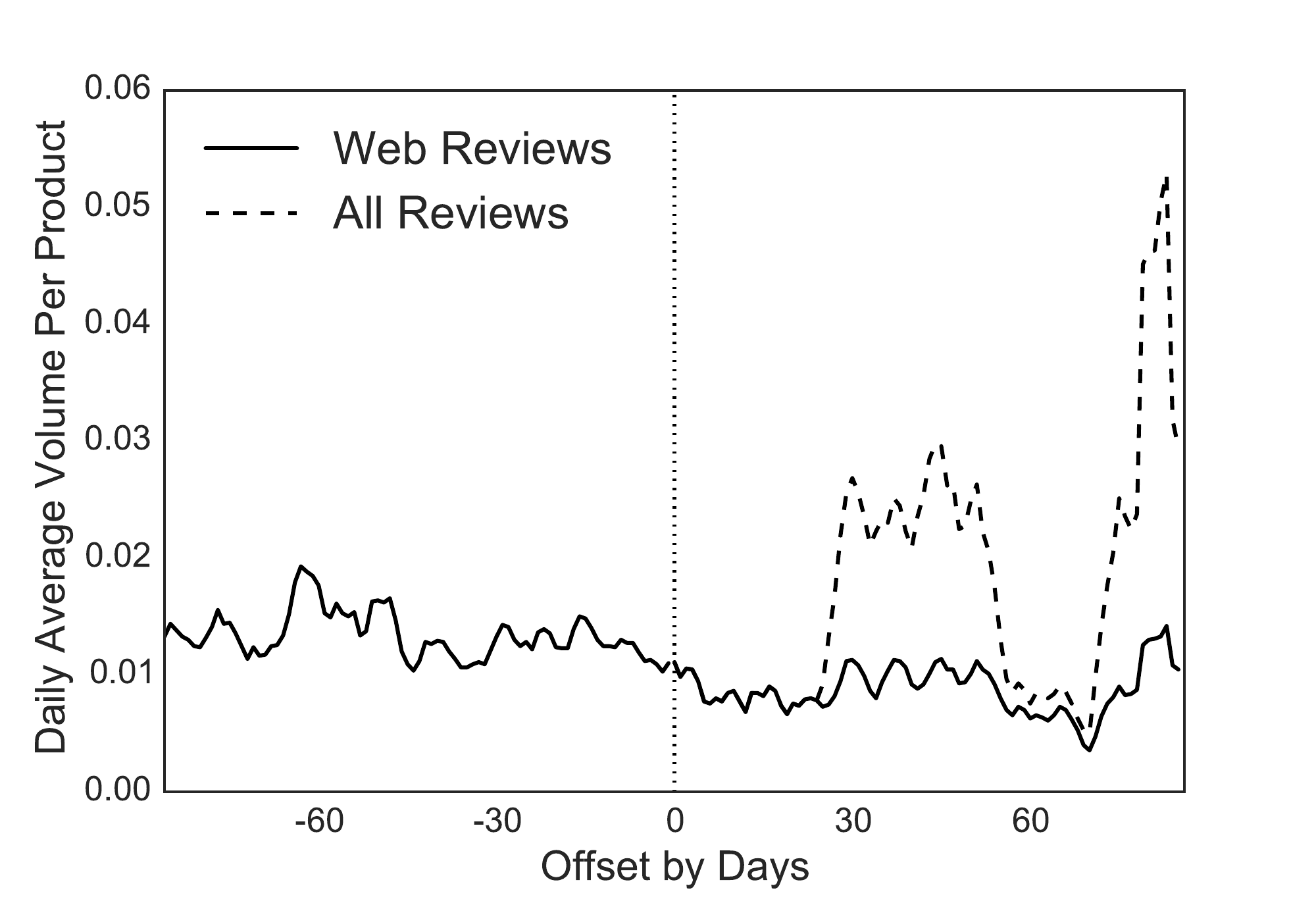}
\caption{Daily average volume per product before and after the introduction of email reviews (dotted vertical line) for web and email reviews.}
\label{volume_bfr_aft_90}
\end{subfigure}
\centering
\begin{subfigure}{.45\textwidth}
\centering
\includegraphics[width=\textwidth]{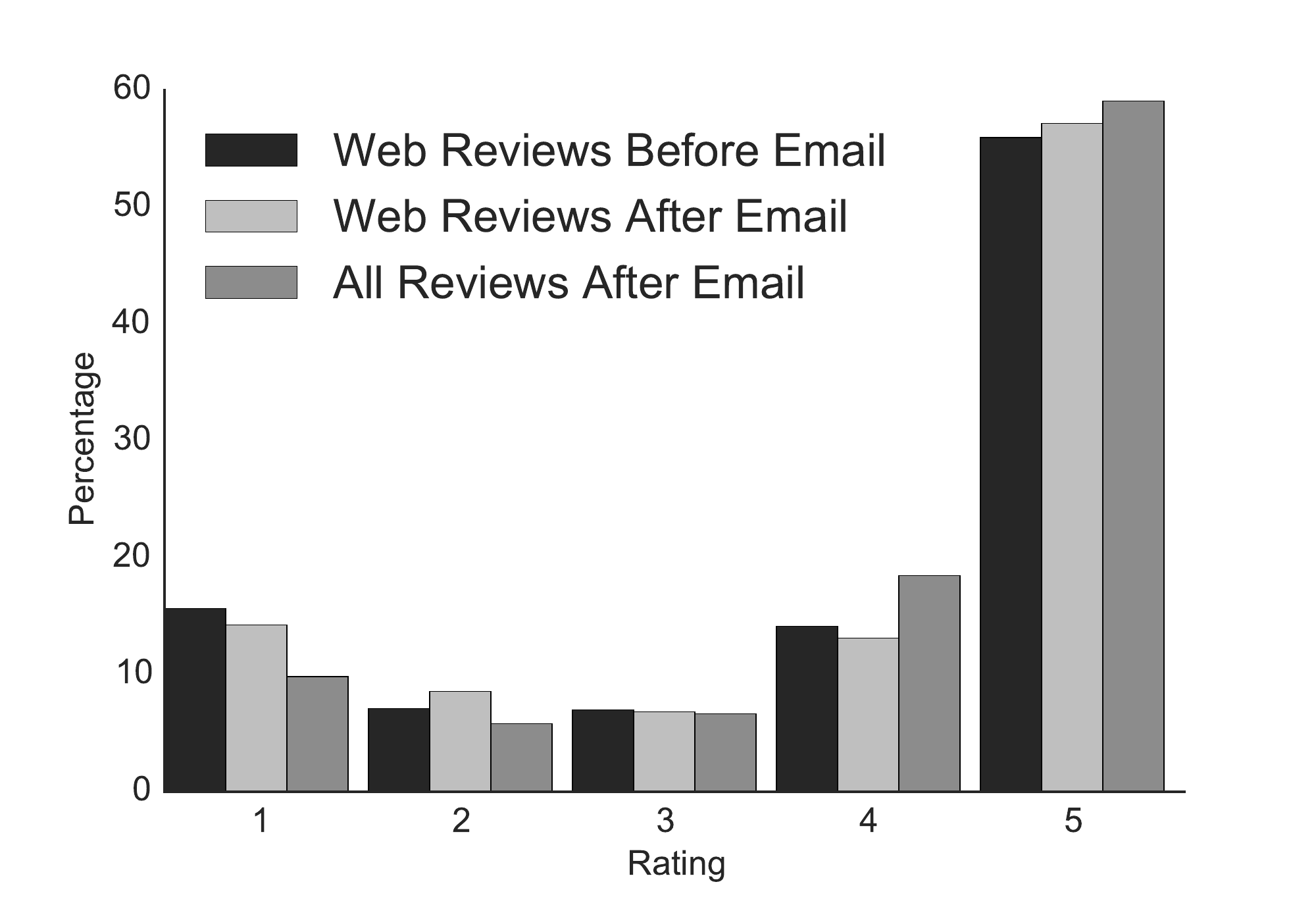}
\caption{Distribution of ratings before and after the introduction of email reviews for web and email reviews}
\label{bfr_after_email}
\end{subfigure}

\caption{Volume and Review Rating before and after the introduction of email reviews}
\label{email_effect_plot}
\end{figure}

 We start with an exploratory look into the data. First, we are interested in understanding the effect in the volume of submitted reviews and we expect to see an increase. Indeed, we find that in the 90 days following the introduction of the email promptings, the volume of reviews increased from around 30 reviews per day to 38 reviews per day. Figure \ref{volume_bfr_aft_90} shows the evolution of the volume of reviews per day for the 90 days before the prompts started and the 90 days after. We see an overall increase in volume, but we also see a decrease in the volume of web reviews submitted, from 30 reviews per day to 19. One potential explanation for this decrease in the volume of web reviews could be that the email promptings cannibalized the web reviews, i.e., they redirected users that would have submitted a review anyway. Another explanation could be that the decline in the volume of web reviews would have happened anyway because of temporal or other (unrelated to the introduction of the promptings) reasons. In order to get a clear picture, we use our control dataset of reviews. We find that the decrease in the volume of the reviews in the treatment year should have been expected even without the introduction of the email promptings, hence we find no evidence of cannibalization.
\par We also explore the effect of the promptings on the characteristics of the submitted reviews. From our findings in Section~\ref{differences}, which showed that email reviews carry, in general, higher ratings than web reviews, we expect email reviews to raise the overall average ratings, and indeed we observe an immediate impact from the promptings: the average rating increases from 3.8 to 4.1. Moreover, as can be seen in Figure \ref{bfr_after_email} there does not seem to be any substantial disturbance in the distribution of web reviews; the average web rating remains roughly stable at 3.8.
\par Finally, we explore the effect on the review text as well. From our results in Section \ref{differences} we know that email reviews are on average shorter, hence we expect their introduction to decrease the average review text length. Indeed, we see a decrease from around 242 characters to 195. 

\subsection{Econometric Model}
In this section we estimate the following model in order to provide statistical rigor to our exploratory results reported above.
\begin{equation}\label{model_2}
y=\alpha_0 + \beta_1\mathrm{treat}+\beta_2\mathrm{post} + \beta_3\mathrm{treat}\cdot\mathrm{post}+e.
\end{equation}
The treat variable indicates if the review is in the treatment or control year. The post variable indicates if the review was submitted after the treatment, i.e., in the last three months of the studied period. The post$\cdot $treat variable represents the interaction of the treat and post variables and, mathematically, it estimates the variation in the treatment group that is {\it not explained} by the control group. The error term is $e$.
\par We are interested in two questions: 1) What was the effect of the email promptings on the entire review platform of the retailer, and, in particular, 2) What was the effect of the email promptings on the existing reviewing population of the retailer, i.e., the web reviews. Model \ref{model_2} can be used to answer both of the above questions. Table \ref{web_and_email} summarizes the results when we estimate Model \ref{model_2} on the entire treatment and control datasets and Table \ref{web_only} summarizes the results when we estimate Model \ref{model_2} while restricting our attention only on web reviews (treatment and control).

\begin{table*}[h!]
\begin{center}
\vspace{0.5cm}
\begin{tabular}{l c c c c  c} 
 \toprule
{\it Dependent Variable}					&Intercept 		&    treat 		&  post &  treat$\cdot$post \\[6pt]
\midrule

\multirow{2}{*}{Volume} 				&  $0.015^{***}$		& $-0.0014$		&$-0.0034^{*}$ 		& $0.075^{***}$\\[6pt]
											&(0.001)						&(0.001)						&(0.01)			&(0.002)		&\\[6pt]
\midrule
\multirow{2}{*}{Review Rating} 		&  $4.05^{***}$		& $-0.19^{***}$		&-0.02				& $0.27^{***}$ \\[6pt]
												&(0.015)				&(0.03)					&(0.02)				&(0.043)		&			\\[6pt]

\midrule

\multirow{2}{*}{\small Log Review Length} &   $5.29^{***}$	& $0.08^{***}$		&-0.003 		& $-0.3^{***}$		\\[6pt]
														&(0.007)				&(0.016)					&(0.01)		&(0.02)		&\\[6pt]

\bottomrule
\end{tabular}
\end{center}
Values in parentheses are standard errors.\\
\small{$^{*}:p<0.05$, $^{**}: p<0.01$, $^{***}:p<0.001$}
\caption{Effect of the email reviews introduction on the entire review ecosystem}
\label{web_and_email}
\end{table*}

\begin{table*}[h!]
\begin{center}
\vspace{0.5cm}
\begin{tabular}{l c c c c  c} 
 \toprule
{\it Dependent Variable}				&Intercept 				&  treat 		&  post	 &  treat$\cdot$post \\[6pt]
\midrule

\multirow{2}{*}{Volume} 			&  $0.015^{***}$			& $-0.0023^{***}$			&$-0.0036^{***}$ 		& -0.0009\\[6pt]
											&(0)							&(0)						&(0)						&(0.001)		&\\[6pt]

\midrule
\multirow{2}{*}{Review Rating} 	&   $4.05^{***}$			& $-0.2^{***}$		&-0.023			& 0.09 \\[6pt]
											&(0.15)						&(0.03)				&(0.022)			&(0.05)		&			\\[6pt]

\midrule

\multirow{2}{*}{\small Log Review Length} 		&  $5.3^{***}$	& $0.082^{***}$		&-0.003 		& 0.04		\\[6pt]
															&(0.007)				&(0.015)				&(0.01)		&(0.024)		&\\[6pt]




\bottomrule
\end{tabular}
\end{center}
Values in parentheses are standard errors.\\
\small{$^{*}:p<0.05$, $^{**}: p<0.01$, $^{***}:p<0.001$}
\caption{Effect of the email reviews introduction on the web reviews}
\label{web_only}
\end{table*}

\subsection{Results} 
We discuss here the results of the model estimations.

\paragraph{Volume} 
\par Figure \ref{volume_bfr_aft_90} showed an increase in the volume of reviews overall but a decrease in the volume of web reviews. The first row of Table~\ref{web_and_email} displays the results of the estimation of Model~\ref{model_2} with volume as the dependent variable using the entire set of reviews (i.e., web and email). The highly significant and positive coefficient for post$\cdot$treat confirms that indeed the overall increase in volume is statistically significant. When we estimate Model~\ref{model_2} while restricting our data only on the web reviews, we see that there is no observable effect on the volume of web reviews from the introduction of email reviews (first row of Table \ref{web_only}). Hence, there is no evidence of cannibalization and the decline in the volume of web reviews should have been expected anyway, even without the introduction of email reviews. 
\par These findings suggest that by sending email promptings, a retailer can incentivize an entirely new segment of their purchasing customer base to submit a review, {\it without affecting the behavior of web reviewers}. This means, that the new overall set of reviews can be more representative.

\paragraph{Rating} Having established that the volume of web reviews is unaffected by the introduction of the email reviews, it's interesting to see if there is any change in the actual {\it reviews} that the web reviewers submit. Our exploratory analysis suggests that there is an increase in the ratings overall but no effect in the ratings coming from web reviews. The second row of Tables~\ref{web_and_email} and \ref{web_only} confirm both of these results. The coefficient for post$\cdot$treat is highly significant and positive when Model~\ref{model_2} is estimated on the entire dataset, which indicates an overall increase in star rating. In contrast, the coefficient for the same variable is not significantly different from zero when we estimate Model~\ref{model_2} only on web reviews, which indicates no change in their star rating. Hence, by sending email promptings, a retailer can access a population that submits, on average, higher ratings without affecting the ratings of the web reviewers.
\paragraph{Review length}
In addition to review rating, we want to see the effect that the email reviews had on the review length overall and on the review length of web reviews in particular. Our exploratory analysis showed a decrease on the average review text length and the coefficient for treat$\cdot$post, shown in the third row of Table \ref{web_and_email}, is negative and statistically significant, hence confirming the result. In contrast, the coefficient for the same variable in the model estimation based only on web reviews, shown in Table \ref{web_only}, is not statistically different from zero, indicating that even though the email reviews had a negative effect on the length of reviews overall, they had no effect on the length of web reviews. 
\par Our overall findings in this section suggest that the email promptings incentivized an entirely new segment of the purchasing population to submit a review, without having any effect on the volume of existing reviewers or the characteristics of their submitted reviews.

\section{Insights and Conclusion}
\par As far as we are aware, our work is the first to compare two fundamentally different sets of reviews for the same products, at the same platform during the same time period. Much existing work has focused on product reviews from platforms like Amazon and Yelp, both platforms that are populated almost entirely by self-motivated reviews. Using our dataset, we are able to replicate many findings that have appeared in that literature, such as the downward trend \citep{godes2012sequential} and bi-modularity \citep{hu2009overcoming} of reviews. But our dataset, also provides insights on a different set of reviews; the reviews that come as a result of an email prompt. We find that these email reviews carry consistently higher ratings than web reviews and do not exhibit any temporal trends. 
\par Furthermore, the introduction of email prompts does not disturb in any way the existing reviewing population while it incentivizes an entirely new segment of the population to submit a review. We think that this finding should provide motivation to retailers to send email prompts to their verified buyers. The reviews overall will become more representative (since a larger segment of the population will be reviewing), more credible (since the new segment of the population that starts reviewing are all verified buyers) and the ratings overall will increase (since the email ratings are on average higher than web ratings). Recent work \citep{zhang2014examining} has shown that source credibility and quantity of reviews can have positive effects on the user's purchase intention. Since the introduction of email prompts will increase the number of submitted reviews and each such submission will be from a `Verified Buyer', the work by \cite{zhang2014examining} suggests that the retailer should expect to see substantial economic benefits from such an introduction.
\par Beyond the review collection, online retailers must also make decisions about how they display the reviews to their users. Currently, most online retailers provide the ability for users to sort the existing reviews by recency, star rating or helpful score. We expect that the default options each retailer implements can have substantial effects on the purchasing decisions of users. Here, we propose an additional policy for retailers to consider: by default, display the average star rating based only on reviews submitted by verified buyers. Since one of the main ways that such verification can happen is by responding to an email prompt from the retailer, these verified reviews will comprise mainly by email reviews that are less bi-modular and stable over time. Hence, they can be a more accurate and credible representation of the true underlying quality of the reviewed product. Moreover, they can be perceived by the reading user as from a more credible source (since they will be submitted by verified buyers), which can have positive effects on the user's purchasing intent \citep{zhang2014examining}. Furthermore, this approach can de-incentivize fake reviews (from unverified buyers), a problem that is becoming increasingly prominent \citep{hu2011manipulation, hu2012manipulation,luca2015fake, mayzlin2012promotional}.
\section{Future Directions}
\par Having argued about the desirable properties that email reviews have, as well the economic benefits they may bring to the retailer, one may ask how aggressively should a retailer pursue such email reviews? For example, a retailer may send multiple email prompts or offer monetary incentives. We believe more research is needed towards understanding the biases that such strategies can introduce in the selection of the reviewing population as well as their submitted reviews. For example, what types of users will be more likely to respond to various types of monetary or social incentives? Can multiple emails annoy the user into submitting a lower rating than they would have otherwise? Or can monetary incentives positively influence the user's perception of the product (and hence their submitted rating)? Insights towards this direction can help practitioners optimally design their review solicitation process.
\par Another interesting future direction is to understand which set of reviews (web or email) is in more agreement with expert ratings, such as ones provided by the Consumer Report magazine. \cite{de2015navigating} found that average user ratings correlate poorly with Consumer Report scores. Moreover, they found that average user ratings are bad at predicting resale values for products. Interestingly, the dataset analyzed by \cite{de2015navigating} is comprised entirely of reviews from Amazon, a platform where the vast majority of reviews are, in the terminology of this paper, web reviews. Would the performance of the average rating (towards the goal of predicting Consumer Report scores and resale values) be improved if email reviews were taken into account? Could email reviews perform even better if considered alone? Insights here can help us understand which set of reviews (web or email) are a more accurate representation of a product's true quality.


\bibliographystyle{plainnat} 
\bibliography{population_selection}

\begin{thebibliography}{31}
\providecommand{\natexlab}[1]{#1}
\providecommand{\url}[1]{\texttt{#1}}
\expandafter\ifx\csname urlstyle\endcsname\relax
  \providecommand{\doi}[1]{doi: #1}\else
  \providecommand{\doi}{doi: \begingroup \urlstyle{rm}\Url}\fi

\bibitem[Anderson(1998)]{anderson1998customer}
Eugene~W Anderson.
\newblock Customer satisfaction and word of mouth.
\newblock \emph{Journal of Service Research}, 1\penalty0 (1):\penalty0 5--17,
  1998.

\bibitem[Anderson and Magruder(2012)]{anderson2012learning}
Michael Anderson and Jeremy Magruder.
\newblock Learning from the crowd: Regression discontinuity estimates of the
  effects of an online review database.
\newblock \emph{The Economic Journal}, 122\penalty0 (563):\penalty0 957--989,
  2012.

\bibitem[Bakhshi et~al.(2014)Bakhshi, Kanuparthy, and Shamma]{bakhshi2014if}
Saeideh Bakhshi, Partha Kanuparthy, and David~A Shamma.
\newblock If it is funny, it is mean: Understanding social perceptions of yelp
  online reviews.
\newblock In \emph{Proceedings of the 18th International Conference on
  Supporting Group Work}, pages 46--52. ACM, 2014.

\bibitem[Cabral and Hortacsu(2010)]{cabral2010dynamics}
Luis Cabral and Ali Hortacsu.
\newblock The dynamics of seller reputation: Evidence from {eBay}.
\newblock \emph{The Journal of Industrial Economics}, 58\penalty0 (1):\penalty0
  54--78, 2010.

\bibitem[Cerasoli et~al.(2014)Cerasoli, Nicklin, and
  Ford]{cerasoli2014intrinsic}
Christopher~P Cerasoli, Jessica~M Nicklin, and Michael~T Ford.
\newblock Intrinsic motivation and extrinsic incentives jointly predict
  performance: A 40-year meta-analysis.
\newblock \emph{Psychological Bulletin}, 140\penalty0 (4):\penalty0 980, 2014.

\bibitem[Chevalier and Mayzlin(2006)]{chevalier2006effect}
Judith~A Chevalier and Dina Mayzlin.
\newblock The effect of word of mouth on sales: Online book reviews.
\newblock \emph{Journal of Marketing Research}, 43\penalty0 (3):\penalty0
  345--354, 2006.

\bibitem[Cui et~al.(2012)Cui, Lui, and Guo]{cui2012effect}
Geng Cui, Hon-Kwong Lui, and Xiaoning Guo.
\newblock The effect of online consumer reviews on new product sales.
\newblock \emph{International Journal of Electronic Commerce}, 17\penalty0
  (1):\penalty0 39--58, 2012.

\bibitem[De~Langhe et~al.(2015)De~Langhe, Fernbach, and
  Lichtenstein]{de2015navigating}
Bart De~Langhe, Philip~M Fernbach, and Donald~R Lichtenstein.
\newblock Navigating by the stars: Investigating the actual and perceived
  validity of online user ratings.
\newblock \emph{Journal of Consumer Research}, 2015.

\bibitem[Dellarocas et~al.(2005)Dellarocas, Awad, and
  Zhang]{dellarocas2005using}
Chrysanthos Dellarocas, Neveen Awad, and M~Zhang.
\newblock Using online ratings as a proxy of word-of-mouth in motion picture
  revenue forecasting.
\newblock \emph{Available at SSRN: http://ssrn.com/abstract=620821}, 2005.

\bibitem[Dichter(1966)]{dichter1966word}
Ernest Dichter.
\newblock How word-of-mouth advertising works.
\newblock \emph{Harvard Business Review}, 44\penalty0 (6):\penalty0 147--160,
  1966.

\bibitem[Duan et~al.(2008)Duan, Gu, and Whinston]{duan2008online}
Wenjing Duan, Bin Gu, and Andrew~B Whinston.
\newblock Do online reviews matter? {An} empirical investigation of panel data.
\newblock \emph{Decision Support Systems}, 45\penalty0 (4):\penalty0
  1007--1016, 2008.

\bibitem[Engstrom and Forsell(2014)]{engstrom2014demand}
Per Engstrom and Eskil Forsell.
\newblock Demand effects of consumers' stated and revealed preferences.
\newblock \emph{Available at SSRN 2253859}, 2014.

\bibitem[Fornell and Bookstein(1982)]{fornell1982two}
Claes Fornell and Fred~L Bookstein.
\newblock Two structural equation models: {LISREL} and {PLS} applied to
  consumer exit-voice theory.
\newblock \emph{Journal of Marketing Research}, pages 440--452, 1982.

\bibitem[Godes and Silva(2012)]{godes2012sequential}
David Godes and Jos{\'e}~C Silva.
\newblock Sequential and temporal dynamics of online opinion.
\newblock \emph{Marketing Science}, 31\penalty0 (3):\penalty0 448--473, 2012.

\bibitem[Hennig-Thurau et~al.(2004)Hennig-Thurau, Gwinner, Walsh, and
  Gremler]{hennig2004electronic}
Thorsten Hennig-Thurau, Kevin~P Gwinner, Gianfranco Walsh, and Dwayne~D
  Gremler.
\newblock Electronic word-of-mouth via consumer-opinion platforms: What
  motivates consumers to articulate themselves on the internet?
\newblock \emph{Journal of Interactive Marketing}, 18\penalty0 (1):\penalty0
  38--52, 2004.

\bibitem[Houser and Wooders(2006)]{houser2006reputation}
Daniel Houser and John Wooders.
\newblock Reputation in auctions: Theory, and evidence from {eBay}.
\newblock \emph{Journal of Economics \& Management Strategy}, 15\penalty0
  (2):\penalty0 353--369, 2006.

\bibitem[Hu et~al.(2009)Hu, Zhang, and Pavlou]{hu2009overcoming}
Nan Hu, Jie Zhang, and Paul~A Pavlou.
\newblock Overcoming the {J-}shaped distribution of product reviews.
\newblock \emph{Communications of the ACM}, 52\penalty0 (10):\penalty0
  144--147, 2009.

\bibitem[Hu et~al.(2011)Hu, Bose, Gao, and Liu]{hu2011manipulation}
Nan Hu, Indranil Bose, Yunjun Gao, and Ling Liu.
\newblock Manipulation in digital word-of-mouth: A reality check for book
  reviews.
\newblock \emph{Decision Support Systems}, 50\penalty0 (3):\penalty0 627--635,
  2011.

\bibitem[Hu et~al.(2012)Hu, Bose, Koh, and Liu]{hu2012manipulation}
Nan Hu, Indranil Bose, Noi~Sian Koh, and Ling Liu.
\newblock Manipulation of online reviews: An analysis of ratings, readability,
  and sentiments.
\newblock \emph{Decision Support Systems}, 52\penalty0 (3):\penalty0 674--684,
  2012.

\bibitem[Li and Hitt(2008)]{li2008self}
Xinxin Li and Lorin~M Hitt.
\newblock Self-selection and information role of online product reviews.
\newblock \emph{Information Systems Research}, 19\penalty0 (4):\penalty0
  456--474, 2008.

\bibitem[Luca(2011)]{luca2011reviews}
Michael Luca.
\newblock {Reviews, reputation, and revenue: The case of Yelp.com}.
\newblock \emph{Harvard Business School NOM Unit Working Paper No. 12-016.
  Available at SSRN: http://ssrn.com/abstract=1928601}, 2011.

\bibitem[Luca and Zervas(2015)]{luca2015fake}
Michael Luca and Georgios Zervas.
\newblock Fake it till you make it: Reputation, competition, and yelp review
  fraud.
\newblock \emph{Harvard Business School NOM Unit Working Paper No. 14-006.
  Available at SSRN: http://ssrn.com/abstract=2293164}, \penalty0 (14-006),
  2015.

\bibitem[Mayzlin et~al.(2012)Mayzlin, Dover, and
  Chevalier]{mayzlin2012promotional}
Dina Mayzlin, Yaniv Dover, and Judith~A Chevalier.
\newblock Promotional reviews: An empirical investigation of online review
  manipulation.
\newblock Technical report, National Bureau of Economic Research, 2012.

\bibitem[Muchnik et~al.(2013)Muchnik, Aral, and Taylor]{muchnik2013social}
Lev Muchnik, Sinan Aral, and Sean~J Taylor.
\newblock Social influence bias: A randomized experiment.
\newblock \emph{Science}, 341\penalty0 (6146):\penalty0 647--651, 2013.

\bibitem[Nagle and Riedl(2014)]{nagle2014online}
Frank Nagle and Christoph Riedl.
\newblock Online word of mouth and product quality disagreement.
\newblock In \emph{Academy of Management Proceedings}, volume 2014, page 15681.
  Academy of Management, 2014.

\bibitem[Resnick et~al.(2006)Resnick, Zeckhauser, Swanson, and
  Lockwood]{resnick2006value}
Paul Resnick, Richard Zeckhauser, John Swanson, and Kate Lockwood.
\newblock The value of reputation on {eBay}: A controlled experiment.
\newblock \emph{Experimental Economics}, 9\penalty0 (2):\penalty0 79--101,
  2006.

\bibitem[Salehan and Kim(2016)]{salehan2016predicting}
Mohammad Salehan and Dan~J Kim.
\newblock Predicting the performance of online consumer reviews: A sentiment
  mining approach to big data analytics.
\newblock \emph{Decision Support Systems}, 81:\penalty0 30--40, 2016.

\bibitem[Salganik et~al.(2006)Salganik, Dodds, and
  Watts]{salganik2006experimental}
Matthew~J Salganik, Peter~Sheridan Dodds, and Duncan~J Watts.
\newblock Experimental study of inequality and unpredictability in an
  artificial cultural market.
\newblock \emph{Science}, 311\penalty0 (5762):\penalty0 854--856, 2006.

\bibitem[Singh(1988)]{singh1988consumer}
Jagdip Singh.
\newblock Consumer complaint intentions and behavior: definitional and
  taxonomical issues.
\newblock \emph{The Journal of Marketing}, pages 93--107, 1988.

\bibitem[Westbrook(1987)]{westbrook1987product}
Robert~A Westbrook.
\newblock Product/consumption-based affective responses and postpurchase
  processes.
\newblock \emph{Journal of Marketing Research}, pages 258--270, 1987.

\bibitem[Zhang et~al.(2014)Zhang, Zhao, Cheung, and Lee]{zhang2014examining}
Kem~ZK Zhang, Sesia~J Zhao, Christy~MK Cheung, and Matthew~KO Lee.
\newblock Examining the influence of online reviews on consumers'
  decision-making: A heuristic--systematic model.
\newblock \emph{Decision Support Systems}, 67:\penalty0 78--89, 2014.

\end{thebibliography}


\end{document}